\documentclass[11pt]{article}
\usepackage{latexsym}
\usepackage{graphics}
\usepackage{times,mathptm}
 \newcommand{\pdf}[0]{{\it pdf}}
 \newcommand{\pdfs}[0]{{\it pdf}s}

\begin{document}
\title {{\Large
   Single or combined location measurements of the same parameter \\
   without prior probability}}
\author {George Kahrimanis \thanks{{\normalsize {\tt anakreon@hol.gr}}}}
\date{{13 May 2003}}
\maketitle

\begin{abstract} {\normalsize
In default of prior probability
the parameter is not defined as a random variable,
hence there can be no genuine prior-free parametric inference.
Nevertheless prior-free predictive inference regarding any future data
is generated directly from the datum of a location measurement.
Such inference turns out as if obtained from a certain \pdf\ 
(``fiducial'') indirectly associated with the parameter.
This false \pdf\ can expedite predictive inference,
but is inappropriate in the analysis of combined measurements
(unless they all are location measurements of the same parameter).
Also it has the same distribution
as the ostensible Bayesian posterior from a uniform ``prior''.
However, if any of these spurious entities is admitted in the analysis,
inconsistent results follow.
When we combine measurements,
we find that the quantisation errors, inevitable in data recording,
must be taken into consideration.
These errors cannot be folded into predictive inference in an exact sense;
that is, we cannot render a predictive distribution of a future datum
except as an approximation. \\ \\
{\bf Keywords}: location measurement; combination of observations;
parametric inference; predictive inference; prior-free inference;
quantisation error; digitisation; frequentist interpretation;
pivotal inference; the fiducial argument; fiducial probability;
intuitive assessment; prior-free assessment.
}\end{abstract}

\section{Introduction}
This work aims at a critical foundation of ``pivotal inference'',
with a generalisation to combined location measurements.
The issue is not in the application,
because the solution is obvious anyway, at least in form,
from direct intuition;
the issue is in stating this solution in non-problematic terms,
within the calculus of probabilities, without any supplementary principle.
Addressing the general problem of prior-free parametric inference
hinges on our understanding this plain situation.

If in some measurement the error is understood as
a random effect influenced only by factors
outside the scope of this measurement, we call it a ``location'' measurement.
For example when we measure the angular distance of two stars in a clear sky,
the probability density function of the error depends
on the properties of the instrument
but is regarded as causally independent of either the true angular distance
or of the resulting datum.

If there is prior probability associated with the location parameter
we apply Bayes' Theorem to generate statistical inference.
The present work deals mainly
with cases in which there is no prior probability.
If we do not know in what way the two stars in the example have been picked,
then we have such a case.

Intuitive interpretation of the result of a location measurement
is quite plain.
If the error is known to follow, for example, a normal distribution,
intuition says that the information we gain about the measured parameter
is expressed by a normal distribution which has the same variance.
Yet against this plausible claim a formal objection is raised, that
either there is prior probability or the parameter is not a random variable;
if the parameter is converted into a random variable
in the course of the analysis,
some paradoxical (at least) consequences follow.
At any rate, let us call this interpretation 
``intuitive'' or ``prior-free'' parametric ``assessment''
(reserving the term `inference' for procedures defined
only within the calculus of probabilities).
It is not distinct from so-called pivotal inference or the fiducial argument.

Related theoretical discussions often involve such terms
that do not provide a clear correspondence to an experimentalist's intuition.
Moreover the logical standing of these probability distributions
has been challenged with regard to certain paradoxes.
To avoid logical inconsistency
this work begins with a careful analysis of the location measurement,
without any direct reference to specific statistical foundations,
like objective Bayesian, fiducial, structural, or pivotal.

However, with the kind of measurements considered here,
we shall encounter no difficulty
in deriving predictive inference regarding any future datum
that is related to the same parameter,
without resorting to parametric inference.
That is, knowing the datum $x_{\,1}$ of the location measurement
and the model ${\Pr}_X(x|\theta)$ of any future measurement,
we obtain directly ${\Pr}_X(x|x_{\,1})$.
Then it turns out that intuition is vindicated somehow,
because predictive inference
seems as if generated from the density function of intuitive assessment.
Yet the latter is only a useful formula,
not genuine probability pertaining to the measured parameter.
In particular this supposed \pdf\ is not meant to be the prior
in a Bayesian analysis of the second measurement --
this and related issues will occupy us in Sec.~\ref{sec:paradoxa} --
though such presumption is harmless
when we combine location measurements of the same parameter.
Let us leave for another work the question whether
predictive inference can in general
be formulated as if from coherent parametric inference,
that is, even when it is based neither on a location measurement
nor on combination of location measurements of the same parameter.

If one assumes an unnormalised uniform density
as prior probability for the parameter
(that is, an unwarranted and possibly improper flat ``prior'')
the related posterior \pdf\ coincides
with the supposed \pdf\ of intuitive assessment.
Moreover a remarkable multitude of proposed statistical foundations
provide the same result to the problem of location measurement
(regarding those, see \cite{Kass} and \cite{Brunero}).
In contrast to them
the present work does not introduce any special axiom
beside the usual postulates of probability.
For that matter, one need not abandon the frequentist interpretation.

\section{A single location measurement}
\subsection{Preliminary analysis}
\subsubsection{Basic considerations}
{\bf Definition.} If a measurement provides a single datum $x$ for estimating
a scalar parameter $\theta$,
the terms `location parameter' and `location measurement' mean that,
in this parametrization, the measurement error, $e$ $\equiv$ $x$$-$$\theta$,
is additive noise of known \pdf\ 
${\Pr}_E(e|{\mathcal B})$ $=$ $f(e)$,
where `${\mathcal B}$' stands for the background evidence or assumptions
that establish this property.
Random variable $E$ is regarded
as causally unrelated to anything else in this model.
The symbol $\theta$ represents any conceivable value of the parameter,
not just the unknown true value.
Each time we simulate the measurement for testing purposes, 
the parameter $\theta$ is assigned some arbitrary value.

We assume that our first measurement is such a measurement,
on the basis of assumptions ${\mathcal B}_{\,1}$.
With reference to any particular value (real or hypothetical) of the parameter,
before the measurement the first datum corresponds to a random variable:
\begin{equation} \label{eq:location}
{\Pr}_{X_{\,1}}(x_{\,1} | {\mathcal B}_{\,1}, \theta) = 
{\Pr}_{E_{\,1}}(x_{\,1}{-}\theta | {\mathcal B}_{\,1}, \theta)
= f_{\,1}(x_{\,1}{-}\theta) \,.
\end{equation}
(Density functions are represented by functions in the stated parametrization.)
If the parameter is unknown, then this \pdf\ is unavailable.

\subsubsection{Bayesian inference for the parameter and for the error}
\label{sec:Bayesian}
It may happen that, independently of the result of the measurement,
we also possess some information ${\mathcal H}$ leading to
the definition of a random variable $\Theta$ corresponding to the parameter,
so that ${\Pr}_{\Theta}(\theta | {\mathcal H})$ $=$ $\pi(\theta)$.
(For example, suppose that the true value of the parameter
is a random sample drawn from a known population.)
Then, using that \pdf\ as the prior,
Bayes' Theorem furnishes the posterior \pdf\ of $\Theta$:
$\; {\Pr}_{\Theta}(\theta | {\mathcal H}, {\mathcal B}_{\,1}, x_{\,1})$
$\;\propto\;$ $\pi(\theta) \, f_{\,1}(x_{\,1}$$-$$\theta) \,$.

From the combination of ${\Pr}_{\Theta}(\theta | {\mathcal H})$
with Eq.~\ref{eq:location},
the pre-measurement joint \pdf\ of $\Theta$ and $X$ is defined,
$\pi(\theta)$ $f_{\,1}(x_{\,1}{-}\theta)$.
The marginal \pdf\ of $E_{\,1}$ ($\equiv X_{\,1}$$-$$\Theta$) is
$f_{\,1}(e_{\,1})$ of course.
After the measurement,
$e_{\,1}$ corresponds to the random variable $x_{\,1}$$-$$\Theta$.
Therefore the \pdf\ of $E_{\,1}$ after the measurement is
\begin{equation}
\label{eq:pdfE}
{\Pr}_{E_{\,1}}(e_{\,1} | {\mathcal H}, {\mathcal B}_{\,1}, x_{\,1})
= {\Pr}_{\Theta}(x_{\,1}{-}e_{\,1} | {\mathcal H}, {\mathcal B}_{\,1}, x_{\,1})
\propto \pi(x_{\,1}{-}e_{\,1}) \, f_{\,1}(e_{\,1}) \,.
\end{equation}

Note that this updated \pdf\ of $E_{\,1}$ can also be viewed
as the posterior of an alternate Bayesian estimation,
concerning $E_{\,1}$ rather than $\Theta$.
An instrument maker is primarily interested in the error of this measurement.
Based on trials or on currently accepted theory,
he has established a prior \pdf\ 
${\Pr}_{E_{\,1}}(e_{\,1} | {\mathcal B}_{\,1})$ $=$ $f_{\,1}(e_{\,1})$
about the error, and is ready to update this \pdf\ 
(with regard to the particular instance only)
when he obtains datum $x_{\,1}$ and prior $\pi(\theta)$ for $\Theta$.
The likelihood of $e_{\,1}$ would be
${\Pr}_{X_{\,1}}(x_{\,1} | {\mathcal H}, e_{\,1})$ $=$ 
${\Pr}_{\Theta}(x_{\,1}$$-$$e_{\,1}| {\mathcal H}, e_{\,1})$
$=$ $\pi(x_{\,1}$$-$$e_{\,1})$.
The resulting posterior for the error would be
$\,\propto\;$ $\pi(x_{\,1}$$-$$e_{\,1})$ $f_{\,1}(e_{\,1})$,
the same as that obtained in Eq.~\ref{eq:pdfE}.
The post-measurement \pdf\ of the error
has been obtained in two different ways
to introduce an argument in the following section.

Frequentist interpretation of the posterior probability of the error
(or any variable)
requires embedding the actual situation
in the most general ensemble of imaginary experiments
that can be conditioned by the same particulars.
In our case, the conditioning consists of datum $x_{\,1}$
in conjunction with prior information ${\mathcal H}$.
Far from being {\it ad hoc}, this embedding schema is implicit
in common and indispensable uses of probability.
For example, after a coin is tossed but before the outcome is seen,
we still associate a probability $1/2$ with `head',
though the outcome is no longer random but only unknown.
Therefore in such cases
we must allow the use of probability to describe one's state of knowledge;
in particular, to weigh forecasts about a fixed unknown quantity,
if in some sense it has been drawn from an ensemble.
In certain other cases, for example
if the unknown quantity is supposed to be a constant of nature,
the notion of probability would be problematic.

\subsubsection{What inference is possible in the absence of prior probability}
\label{sec:prior_free}
Under the assumption that the parameter is a random variable,
Sec.~\ref{sec:Bayesian} contains two equivalent ways of reckoning
the posterior probability of the error.
Based on the rationale of the second derivation,
we can form the following argument.
Suppose that there is no prior probability associated with $\theta$;
then the instrument maker cannot apply Bayes' Theorem
to update his prior for the error;
therefore the \pdf\ of $E_{\,1}$ after the measurement will remain
$f_{\,1}(e_{\,1})$.

The ensemble of imaginary experiments
corresponding to a prior-less location measurement
involves all possible values of the parameter,
though with an indeterminate distribution,
because the parameter is not a random variable.
Without a definite distribution of values for the parameter,
conditioning this ensemble on the datum alone
cannot update the \pdf\ of the error.\footnote
{{\normalsize
We can represent the unconditioned ensemble
by means of a sequence of actual simulated measurements,
each starting with some arbitrary $\theta_{\,i}$,
provided that the distribution of these values is understood as irrelevant.
Then conditioning on the datum alone
(not on the presumed true value of the parameter, which is inaccessible)
cannot modify the distribution of the error,
which remains $f_{\,1}(e_{\,1})$.
}}

Caution must be exercised in interpreting random variable $x_{\,1}$$-$$E_{\,1}$,
after the measurement. (For that matter, before the measurement
$x_{\,1}$ neither is constant nor corresponds to a random variable,
except only in relation to any fixed hypothetical value of the parameter;
that is, before the measurement
there is no available definite distribution associated with the datum.)
It is tempting to identify $x_{\,1}$$-$$E_{\,1}$ with the parameter itself,
since its unknown true value is equal to
the unknown true value of the parameter.\footnote
{{\normalsize
Before the measurement, the expression $x_{\,1}$$-$$E_{\,1}$
is equivalent to $\theta$,
but the meaning changes after the measurement:
$x_{\,1}$ is restricted to the actual result but $E_{\,1}$ is still understood
with reference to the ensemble of all measurements.
}}
Yet this premise would lead to paradoxical, at least, conclusions
(as in Sec.~\ref{sec:paradoxa}); it is not adopted here.
The parameter remains an unknown constant only,
not converted into a random variable.

Now we come to the problem of predictive inference.
In general the second measurement is modeled as
\begin{equation} \label{eq:model}
{\Pr}_X(x | {\mathcal M}, \theta) = p(x;\theta)
\end{equation}
that is, $x|{\mathcal M}, \theta$ corresponds to a random variable.

We can also match $x|{\mathcal M},x_{\,1},e_{\,1}$ to a random variable,
because each possible value $e$
corresponds to $\theta$ $\equiv$ $x_{\,1}$$-$$e_{\,1}$:
\begin{equation} \label{eq:likelihood_e}
{\Pr}_X(x | {\mathcal M},x_{\,1},e_{\,1}) = p(x;x_{\,1}{-}e_{\,1})
\end{equation}
so that, in conjunction with the \pdf\ of $E_{\,1}$,
joint \pdf\ of $X$ and $E_{\,1}$ is defined:
\begin{equation}
{\Pr}_{X,E_{\,1}}(x, e_{\,1} | {\mathcal B}_{\,1},{\mathcal M},x_{\,1})
= p(x;x_{\,1}{-}e_{\,1}) \, f_{\,1}(e_{\,1}) \,.
\end{equation}
Predictive inference for $X$
amounts to the marginal \pdf\ of this variable,\footnote
{{\normalsize
If we take into account that we never know $x_{\,1}$ directly
except by means of some digital transcript $y_{\,1}$,
we shall be interested in conditioning by $y_{\,1}$ rather than by $x_{\,1}$.
This problem is addressed in Sections
\ref{sec:quantisation_error0} and \ref{sec:quantisation_error1}.
}}
\begin{equation} \label{eq:pred_X}
{\Pr}_X(x | {\mathcal B}_{\,1},{\mathcal M},x_{\,1})
= \int de \, p(x;x_{\,1}{-}e_{\,1}) \, f_{\,1}(e_{\,1}) \,.
\end{equation}

We have obtained prior-free predictive inference about any future measurement
directly,
thus bypassing the problem of prior-free parametric inference.
Not only the assumption of a \pdf\ for the parameter is not required,
but it would annul this derivation.
Of course we depend on the assumption that
there is a single, constant true value of the parameter,
otherwise the substitution of $\theta$ (in the model of the second measurement)
with $x_{\,1}$$-$$e_{\,1}$ would not be justified.

The parameter itself is not a random variable,
yet there is a remarkable coincidence.
If we only pretend that
in the future measurement the parameter is a random variable
with the same \pdf\ as $x_{\,1}$$-$$E_{\,1}$
the resulting predictive inference for $X$ will turn out
the same as in Eq.~\ref{eq:pred_X}.
Therefore there is utility
in communicating the \pdf\ of $x_{\,1}$$-$$E_{\,1}$,
because it happens to generate correct predictive inference.
For that matter,
only in this capacity our intuitive parametric assessment
can acquire a proper definition.\footnote
{{\normalsize
Fisher has remarked that
the problem of fiducial predictive inference based on datum $x_{\,1}$
can be solved ``directly'',
not only ``after the [...] distribution of the population parameter[...]
has been obtained'' (\cite{Fisher35}, Sec. II),
``without discussing the possible values of the parameter $\theta$''
(\cite{Fisher56}, Sec. V.3).
Yet he defines fiducial prediction
only as derived from fiducial probability of the parameter values;
consequently the simplification he mentions is only a secondary issue.
In the present work predictive inference is defined
in the absence of any distribution for $\theta$;
therefore the possibility to also calculate it
as if from some intuitive density function of the parameter
is regarded as fortuitous,
proved in the case of location measurements
but not yet guaranteed to be generally true.
}}
\begin{equation} \label{eq:pivotal0}
{\Pr}_{(x_{\,1}{-}E_{\,1})}(\theta | {\mathcal B}_{\,1},x_{\,1})
= {\Pr}_{E_{\,1}}(x_{\,1}{-}\theta | {\mathcal B}_{\,1},x_{\,1})
= f_{\,1}(x_{\,1}{-}\theta).
\end{equation}
The same result, in form, has been obtained also by means of applying
fiducial, pivotal, structural, or objective Bayesian ``inference'',
but in this work we do not consider any kind of inference not based
on the calculus of probabilities exclusively.

The density function of intuitive parametric assessment
(that is, the \pdf\ of $x_{\,1}$$-$$E_{\,1}$)
has been defined
without reference to what the parameter stands for;
it represents only whatever the apparatus has measured.
It makes no difference whether the true value of the parameter
is understood to be, for example, a constant of nature
or some quasi-constant magnitude (like the radius of a star).
The \pdf\ is not {\em of }that magnitude
but only related to it, as an accessory of predictive inference.
Therefore this \pdf\ 
is not in conflict with the frequentist interpretation of probability.

\subsubsection{Essential distinctions and caveats}
Sometimes the issue of prior-free assessment is confused
due to considerations assuming a joint \pdf\ of $\Theta$ and $X_{\,1}$.
For that matter, it cannot be denied that the existence
of some \pdf\ related to the parameter after the measurement,
along with the family of conditional \pdfs\ 
${\Pr}_{X_{\,1}}(x_{\,1} | \theta)$, defines some joint \pdf;
yet this \pdf\ is not of $\Theta$ and $X_{\,1}$
but of $x_{\,1}$$-$$E_{\,1}$ and $X_{\,2}$,
where $X_{\,2}$ represents the datum
of a hypothetical second {\it iid} measurement.
Besides, if one assumed a joint \pdf\
of some $\Theta$ (defined independently) and $X_{\,1}$,
then the Bayesian method would be applicable;
the marginal \pdf\ of that $\Theta$ would be the prior.

If one assumes a prior uniform in the location parameter (we have not done so)
the posterior will be represented by the same density function
as that of prior-free assessment,
though the two functions are different in intent.
In this work, if a prior happens to generate the same posterior as
the \pdf\ of prior-free assessment,
it will be called the ``false prior'' of the measurement
to emphasise that (like a false ceiling or false teeth)
it has a function (it generates the same result as prior-free assessment)
but is not the real thing.\footnote
{{\normalsize
That is, a false prior for $\theta$ is not meant to be combined with
the parametric family ${\Pr}_X(x | \theta)$ 
to produce a joint \pdf\ of $\Theta$ and $X$.
The only use of a density that is known to be a false prior
is to be multiplied with the likelihood, ${\Pr}_X(x | \theta)$,
in order to calculate the density function of intuitive assessment.
}}
In any specific problem, existence of a false prior would simply mean that
prior-free prediction is equivalent to prediction based on that prior
(as it happens; not because Bayes' Theorem applies).
Non-existence of a false prior would imply either
that predictive inference cannot be reduced
to any presumptive \pdf\ of the parameter,
or a situation in which the ratio of that \pdf\ to the likelihood
depends not only on $\theta$ but also on $x$.

The case of a location measurement without prior probability
is so intuitively transparent that it is inconceivable
for any proposed statistical foundation to provide any solution
other than a density function
of the same profile as that of Eq.~\ref{eq:pivotal0},
that is, equivalent to the posterior obtained from a uniform false prior.
``Virtually all default Bayesian methods recommend this conditional prior,
as do various ``structural'' and even frequentist approaches''
(\cite{Brunero}, Sec.~4.3, referring to a more general case).
Therefore, in the case of a location measurement at least,
it seems legitimate for an experimenter to follow plain intuition and
consider the option of no prior.
Of course we must examine in what circumstances this option is appropriate;
on the other hand, we must also assess when
prior probability can be asserted on the basis of educated guessing only.
These questions are left for another work.

\subsubsection{Can Bayesian estimation be combined with prior-free assessment?}
\label{sec:paradoxa}
If someone regards the \pdf\ of prior-free assessment
as converting the parameter into a random variable,
then in the analysis of a second measurement
Bayes' Theorem would be applicable,
with this \pdf\ providing prior probability.
It is not surprising that
this consideration leads to inconsistency in the results.
For instance, if we combine two location measurements
regarding not the same parameter but two functions of the same parameter,
the posterior \pdf\ 
depends on the order of analysing the two measurements.

Yet even if we reckon strictly within the limits of predictive inference,
the same predicament seems to arise.
When the second datum, $x$, is obtained,
Eq.~\ref{eq:likelihood_e} provides
the likelihood of any value $e_{\,1}$ of the error;
the likelihood can be combined with the prior \pdf, $f_{\,1}(e_{\,1})$,
to form the posterior \pdf\ of $E_{\,1}$.
{\em If this is correct}, then
new \pdf\ of $x_{\,1}$$-$$E_{\,1}$ will be obtained,
which in turn may be used
for predictive inference regarding a third datum, and so on.
In effect random variable $x_{\,1}$$-$$E_{\,1}$,
with updated \pdf\ after each measurement,
would act as a surrogate random variable for the parameter,
even after we combine two or more location measurements.
Yet in this way the parameter turns effectively into that random variable,
in spite of the initial assumption.
If this concern seems remote from practical applications,
let us consider the combination of two location measurements
regarding not the same parameter but two functions of the same parameter,
like the diameter and the volume of a sphere.
If we update the \pdf\ of the error in the way suggested above,
then predictive inference concerning any future third measurement
depends on the order of combining the two data
(the method is equivalent to admitting the false prior
of the first measurement).
Unless we suppose that information may have some metaphysical consequence,
we cannot accept that predictive inference about any third datum
may depend on the order of receiving the first two data.
Almost certainly there is some error in the argument,
and we must find it
(otherwise this inconsistency
would be credited to the common postulates of probability calculus).

\subsubsection{Predictive inference is known only as an approximation}
\label{sec:quantisation_error0}
A careful review locates the fault in overlooking
an approximation attendant to employing Eq.~\ref{eq:likelihood_e}.
The ``analog'' datum $x_{\,1}$ is registered as a numerical record $y_{\,1}$,
using a scale of step $P_{\,1}$,
so that a quantisation error $r_{\,1}$ ($\equiv$ $y_{\,1}$$-$$x_{\,1}$),
however small, is unavoidable,
constrained in the range $[-P_{\,1}/2, P_{\,1}/2)$.
Therefore corresponding to any conceivable value $e_{\,1}$
($\equiv$ $x_{\,1}$$-$$\theta$)
there is no single value $\theta$, but rather some small range of values.

Nevertheless the pursuit of predictive inference is not impaired thereby,
except that we end up with an approximation
of the unknown true density function.
Since $x_{\,1}$ is strictly speaking inaccessible,
Eq.~\ref{eq:pred_X} cannot be our goal.
We can redefine predictive inference,
to be conditioned by $y_{\,1}$ and $r_{\,1}$ instead of $x_{\,1}$:
\begin{equation} \label{eq:pred_X_b}
{\Pr}_X(x | {\mathcal B}_{\,1},{\mathcal M},y_{\,1},r_{\,1})
= \int de_{\,1} \, p(x;y_{\,1}{-}r_{\,1}{-}e_{\,1}) \, f_{\,1}(e_{\,1}) \,.
\end{equation}
Inasmuch as $r_{\,1}$ cannot be known,
we have not yet produced exact predictive inference.
We would like to define
``${\Pr}_X(x | {\mathcal B}_{\,1},{\mathcal M},y_{\,1})$''
but, as we shall see in Sec.~\ref{sec:quantisation_error1},
we cannot apply any exact meaning to this expression
in the absence of prior probability.
Nevertheless for small enough $P_{\,1}$
(we are interested only in the limit of vanishing $P_{\,1}$
because in general we want to keep round-off errors so small
as to be insignificant)
and assuming smoothness of $f_{\,1}(\cdot)$ and $p(\cdot;\cdot)$,
the true but unknown
${\Pr}_X(x | {\mathcal B}_{\,1},{\mathcal M},y_{\,1},r_{\,1})$
is uniformly approximated by the following density function
(the limiting operation assumes an imagined sequence
with $P_{\,1}$ going to $0$;
the corresponding values $r_{\,1}$, in $[-P_{\,1}/2, P_{\,1}/2)$,
are of no consequence in defining this limit)
\begin{equation} \label{eq:pred_X_c}
{\lim}~~{\Pr}_X(x | {\mathcal B}_{\,1},{\mathcal M},y_{\,1},r_{\,1})
\; = \; \int de_{\,1} \, p(x;y_{\,1}{-}e_{\,1}) \, f_{\,1}(e_{\,1}) \,,
\end{equation}
not dependent on $r_{\,1}$.
In practice it takes the place of the non-existent
``${\Pr}_X(x | {\mathcal B}_{\,1},{\mathcal M},y_{\,1})$''.\footnote
{{\normalsize
Of course digitisation steps cannot be made arbitrarily small;
on the contrary,
in any set of measurements, $P_{\,i}$ are fixed, finite magnitudes.
The limiting operation is intended to justify the approximation
of a true but unavailable \pdf\ by a known density function.
}}

Yet in view of this complication
the troubling Bayesian argument in Sec.~\ref{sec:paradoxa}
must be replaced by a more detailed examination,
as in the following sections;
the perplexing inconsistency that it had generated disappears,
only at the cost of our having to reconsider
the method of updating the \pdf\ of $e_{\,1}$.
For that matter, we shall see that $r_{\,1}$ does not correspond
to any exactly defined \pdf,
therefore there is no exactly defined updated distribution for $e_{\,1}$,
unless it is conditioned by the unknown $r_{\,1}$.

\subsection{Quantisation error in a single location measurement
without prior}
\label{sec:quantisation_error1}
Even in the plainest measurement,
every time a numerical record is kept a quantisation error is generated,
also known as ``digitisation'' error.
We are justified to neglect it
in the case of a single location measurement,
because it is possible to define the \pdf\ of prior-free assessment
with good approximation without reference to this error.
Yet combinations of measurements cannot be studied
if the quantisation error is neglected.
To prepare that study, we include now
the quantisation error even in the analysis of a single location measurement.

For simplicity, let us assume that the designer takes sufficient care so that
digitisation does not involve any error beside the quantisation error,
or rather that such errors are negligible
in comparison with the principal ``analog'' error, $E$.
The quantisation error remains for ever unknown
(unless we preserve and digitise again the same $x$)
unlike the total error, $e$$+$$r$,
which can be estimated to any desired precision,
with subsequent measurements.

Note that need for consideration of the quantisation error
does not arise only in prior-free inference.
Without it, even the model problem
of Bayesian inference with continuous sample space
would be impossible to solve
because the probability of any real datum $x$ is zero.
Actually the likelihood of $\theta$ is defined
not for some datum $x$ but for a range of $x$
between $y{-}$$P/2$ and $y{+}$$P/2$.
The usual underlying assumption is
that the assessment of $P$ does not involve $\theta$.
It is not required that $P$ be very small;
smallness of $P$ only simplifies the calculations.

Though it is commonly assumed that if $P$ is small enough
then $r$ is distributed approximately uniformly over its range,
here we must examine why and in what sense this is valid.

\begin{figure}[h]
\begin{center}
\includegraphics{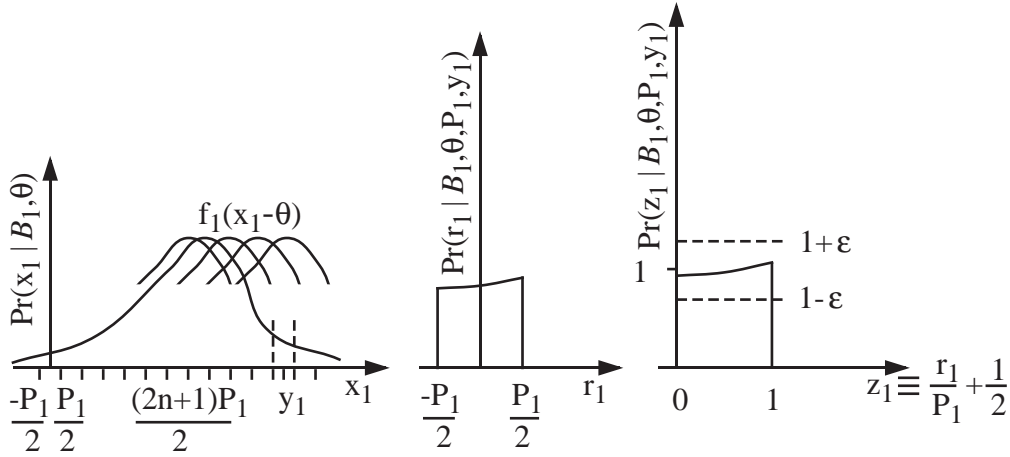}
\caption
{\label{fig:digit}
The result of a digitisation
corresponds to an interval centered at $y_{\,1}$, of width $P_{\,1}$.
The ``true'' distribution of quantisation error $r_{\,1}$
(imagining an ensemble of measurements
starting with $\theta$ and ending with $y_{\,1}$)
is $\propto$ $f_{\,1}(y_{\,1}$$-$$r_{\,1}$$-$$\theta)$.
Inasmuch as $\theta$ is unknown and not a r.~v.,
this distribution is indefinite.
Yet in some sense it can be approximated by a rectangular distribution,
regardless of $\theta$,
though at the cost of excluding from the analysis
all $\theta$ whose likelihood is smaller than some threshold.}
\end{center}
\end{figure}

We must clarify in what sense we assign probability density to $r_{\,1}$
after the measurement.
Typically the quantisation error of a digitisation remains unknown,
but for the sake of this argument now we shall imagine that we measure it.
Let us assume a second digitisation of the same analog quantity $x_{\,1}$,
with a much smaller quantisation step, so that we can estimate $r_{\,1}$.
The ``true'' distribution of $r_{\,1}$ is understood in reference to
an ensemble of {\it iid }measurements
of the same parameter, each followed by the two digitisations,
the first of which happens to produce the same $y_{\,1}$
(else it is excluded from this ensemble).

Thus specified, the distribution of $r_{\,1}$ is defined
in terms of the distribution of $x_{\,1}|{\mathcal B}_{\,1},\theta$
which is $f_{\,1}(x_{\,1}$$-$$\theta)$,
from Eq.~\ref{eq:location}.
(That is, the distribution of $r_{\,1}$ in $[-P_{\,1}/2, P_{\,1}/2)$
corresponds to a normalised slice of $f_{\,1}(x_{\,1}$$-$$\theta)$,
with $x_{\,1}$ between
$\, y_{\,1}$$-$$P_{\,1}/2 \,$ and $y_{\,1}$$+$$P_{\,1}/2$ --
see Fig.~\ref{fig:digit}.)
\begin{equation}
{\Pr}_{R_{\,1}}(r_{\,1}|{\mathcal B}_{\,1},\theta,y_{\,1})
\propto f_{\,1}(y_{\,1}{-}r_{\,1}{-}\theta) \,.
\end{equation}
However this definition alone
cannot amount to any determinate \pdf\ being assigned to $r_{\,1}$,
because the true $\theta$ is unknown.\footnote
{\label{n:posterior_R} {\normalsize 
If there is a prior for $\theta$,
we usually are not concerend with this true but elusive \pdf\ 
because then $\theta$ is a random sample from a population,
so that we can fold
the posterior \pdf\ of $\Theta$ with $f_{\,1}(y_{\,1}$$-$$r_{\,1}$$-$$\theta)$
to produce the posterior \pdf\ associated with $r_{\,1}$,
whether $P_{\,1}$ is small or not.
The approximation of this \pdf\ by a rectangular one for small $P_{\,1}$
is warranted on the basis of smoothness and bounded derivatives.
}}
Operative meaning is generated by the following considerations.

Approximating this unknown distribution by a rectangular one
would be justified
if we could prove that, when $P_{\,1}$ approaches $0$,
the distribution of $r_{\,1}/P_{\,1}$$+$$1/2$ (in $[0,1)$)
converges uniformly to $1$ for all $\theta$.
Whether this is true depends on $f_{\,1}(\cdot)$.
Even in the case of a normal distribution the premise is not true:
considering any $P_{\,1}$-wide slice of a normal distribution,
the log of the ratio of left-to-right heights
is proportional to the distance from the mean;
therefore it goes to $\pm\infty$ for $\theta$ $\rightarrow$ $\pm\infty$.

With a normal $f_{\,1}(\cdot)$,
the only way to contrive this uniform convergence
is by excluding from consideration
all $\theta$ whose likelihood ($f_{\,1}(y_{\,1}{-}\theta)$)
is less than some arbitrary small threshold.
With this provision not only normal distributions but also
any smooth $f_{\,1}(\cdot)$ with bounded derivatives
can be made to fulfil the condition of uniform convergence.
Detailed proof is rather straightforward.
In this way we have defined a provisional approximation
of the post-measurement \pdf\ of $r_{\,1}$
by the rectangular distribution.

Perhaps a summary is helpful here.
To advance our main undertaking,
accurate combination of location measurements without prior probability,
we have seen the necessity of taking into account quantisation errors.
These errors are commonly associated approximately
with rectangular distributions if the step is small enough.
Such association would be justified
only if the true distribution converges uniformly to the rectangular shape
as $P_{\,1}$ goes to $0$.
Yet for the post-measurement \pdf\ of $r_{\,1}$ this is not true in general.
To justify approximating this \pdf\ by a rectangular distribution
not only we assume a small-enough $P_{\,1}$
but also we provisionally exclude all $\theta$
whose likelihood is smaller than some arbitrary threshold.

This provision amounts to setting an arbitrary high confidence level
$1$$-$$\varepsilon$
to generate a (very wide) confidence interval, or group of intervals,
for $\theta$,
using likelihood to provide the requisite ordering.
Then the premise of uniform convergence is at the same confidence level.
Under this provision,
the quantisation error corresponds approximately
to a random variable $R_{\,1}$ of rectangular \pdf\
in $[-P_{\,1}/2, P_{\,1}/2)$,
as if independent of $E$ and without dependence upon $\theta$.

The pre-measurement \pdf\ of $r_{\,1}$ is an average over all $y_{\,1}$
(that is, $0$, $\pm P$, $\pm 2P$, ...) with reference to $\theta$:
\begin{equation} \label{eq:pre_prob_r}
{\Pr}_{R_{\,1}}(r_{\,1}|{\mathcal B},\theta) =
{\sum}_{y_{\,1}} \, {\Pr}_{Y_{\,1}}(y_{\,1}|{\mathcal B}_{\,1},\theta) \,
{\Pr}_{R_{\,1}}(r_{\,1}|{\mathcal B}_{\,1},\theta,y_{\,1}) =
{\sum}_{y_{\,1}} f_{\,1}(y_{\,1}{-}r_{\,1}{-}\theta) \,.
\end{equation}
It depends on $\theta$ in a periodic way, with period $P_{\,1}$.
Like the related post-measurement \pdf, it cannot be known exactly
because $\theta$ is unknown and not a random variable.
Uniform approximation by a rectangular distribution
can be derived from general assumptions of smoothness
in combination with Eq.~\ref{eq:pre_prob_r},
without need for any special provision.

In conclusion, in the limit of small $P_{\,1}$,
the quantisation error corresponds approximately to a random variable $R_{\,1}$
of rectangular distribution, the same before and after the measurement,
if we impose on $\theta$ some threshold (however small) of likelihood.

Since we have established some meaning
for associating $r_{\,1}$, approximately,
with a random variable $R_{\,1}$ of rectangular distribution,
now we may fold this \pdf\ of $R_{\,1}$ with Eq.~\ref{eq:pred_X_b}
to provide approximate meaning
to ``${\Pr}_X(x | {\mathcal B}_{\,1},{\mathcal M},y_{\,1})$''.
Yet this would not be approximation of a defined if unknown \pdf\ 
but only taking advantage of approximations
to define a density function
that is not otherwise meaningful
(inasmuch as there is no prior for $\theta$ -- see n.~\ref{n:posterior_R}).
Well defined predictive inference is provided
by the true though not exactly known
${\Pr}_X(x | {\mathcal B}_{\,1},{\mathcal M},y_{\,1},r_{\,1})$,
approximated in Eq.~\ref{eq:pred_X_c}.

Rather than being regrettable,
the lack of such predictive inference
that would be both accessible and exactly defined
is our way out of the inconsistency mentioned in Sec.~\ref{sec:paradoxa}.
For that matter, uncertainty is the very subject of probability calculus,
so nothing is lost if it is combined with approximation.

The considerations outlined in Sec.~\ref{sec:prior_free}
regarding the utility of the density function of prior-free assessment --
now $f_{\,1}(y_{\,1}$$-$$\theta)$ rather than $f_{\,1}(x_{\,1}$$-$$\theta)$ --
have not been affected.
This density function does not express probability density of $\theta$
but if used in that sense it provides predictive inference,
within approximation depending on the size of the digitisation step.

\section{Combining two location measurements}
\subsection{Statement of the problem}
\label{sec:naive}
So far we have considered one completed location measurement
in relation to a future generic measurement.
Now we consider two possibly different completed location measurements
of the same parameter, in relation to a future generic measurement.
That is, the two errors \\
$e_{\,1}$ $\equiv$ $x_{\,1}$$-$$\theta \;$ and
$\; e_{\,2}$ $\equiv$ $x_{\,2}$$-$$\theta$\\
follow true \pdfs\ independently of any other variable and of each other:\\
${\Pr}_{E_{\,1}}(e_{\,1} | {\mathcal B}_{\,1})$ $=$ $f_{\,1}(e_{\,1}) \;$ and 
$\; {\Pr}_{E_{\,2}}(e_{\,2} | {\mathcal B}_{\,2})$ $=$ $f_{\,2}(e_{\,2})$ ~. \\
Therefore, before the data become known, the joint \pdf\ of the two errors is
\begin{equation}
\label{eq:joint1}
{\Pr}_{E_{\,1},E_{\,2}}(e_{\,1},e_{\,2} | \{{\mathcal B}_{\,i}\}) = f_{\,1}(e_{\,1})f_{\,2}(e_{\,2}).
\end{equation}

When the two data $x_{\,1}$ and $x_{\,2}$ are taken into account,
$e_{\,1}$ and $e_{\,2}$ are jointly constrained by the condition
\begin{equation}
\label{eq:naive}
e_{\,1} - e_{\,2} = x_{\,1} - x_{\,2}
\end{equation}
which defines a straight line $l$ in the $(e_{\,1},e_{\,2})$ plane,
parametrized by $\theta$:
\begin{eqnarray}
\label{eq:param1}
e_{\,1} &=& x_{\,1} - \theta \nonumber \\
e_{\,2} &=& x_{\,2} - \theta .
\end{eqnarray}
Therefore we need define relative probability for segments of this line.
But such probability may be defined
only with reference to some limit,
because joint probability vanishes for line segments.

However, if we compare two segments of $l$ of equal length,
chosen so that, at all points of the first segment, the joint \pdf\ 
is higher than at every point of the other segment,
we intuitively perceive that
the probability of the first segment is higher than that of the second one.
For an expanation why this is so, we must take into account
the errors generated in the digitisation of data.

\subsection{Taking into account the quantisation errors}
\label{sec:quantisation_error2}
In terms of the digitised outcomes $y_{\,i}$ ($i=1,$$2$)
(these are the ones actually issued)
Eq.s~\ref{eq:param1} become
\begin{eqnarray}
e_{\,1} &=& y_{\,1} - r_{\,1} - \theta \nonumber \\
e_{\,2} &=& y_{\,2} - r_{\,2} - \theta.
\label{eq:param2}
\end{eqnarray}
By eliminating $\theta$ we obtain
\begin{equation} \label{eq:keen}
e_{\,1} - e_{\,2} = y_{\,1} - y_{\,2} + r_{\,2} - r_{\,1}
\end{equation}
instead of Eq.~\ref{eq:naive}.
Since each $r_{\,i}$ is constrained within $[-P_{\,i}/2, P_{\,i}/2)$,
the above equation defines
a strip in the $(e_{\,1},e_{\,2})$ plane.

It will be convenient to transform the pair of random variables 
$(E_{\,1},E_{\,2})$ into $(E_+,E_-)$:
\begin{eqnarray} \label{eq:rotation_E}
E_+ &\equiv& (E_{\,1} + E_{\,2})/\sqrt{2} \nonumber \\
E_- &\equiv& (E_{\,2} - E_{\,1})/\sqrt{2} \,.
\end{eqnarray}
This is equivalent to a $\, \pi/4 \,$ rotation of coordinate axes.

The related Jacobian determinant is $1$,
therefore the pre-measurement joint \pdf\ of $E_+$ and $E_-$ is
\begin{equation} \label{eq:joint2}
{\Pr}_{E_+,E_-}(e_+,e_- | \{{\mathcal B}_{\,i}\}) \; = \;
f_{\,1}((e_- {-} e_+)/\sqrt{2}) \; f_{\,2}((e_- {+} e_+)/\sqrt{2}) \,.
\end{equation}

\begin{figure}[h]
\begin{center}
\includegraphics{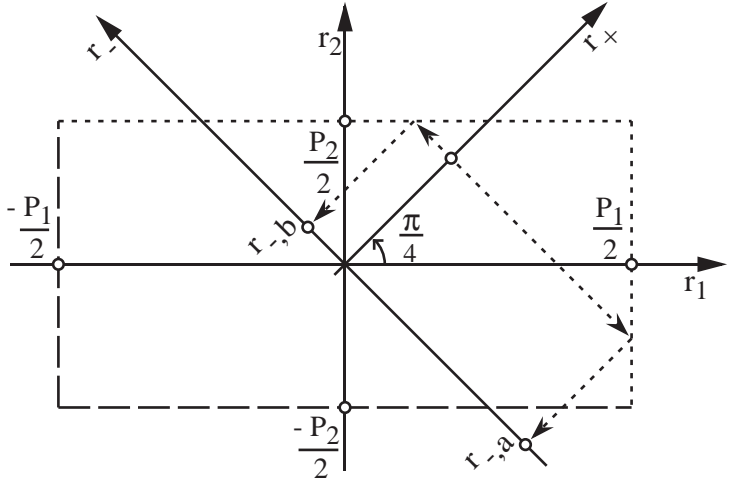}
\caption{\label{fig:r1r2}
Quantisation errors of two measurements}
\end{center}
\end{figure}

A corresponding rotation of axes in the the $(r_{\,1},r_{\,2})$ plane
takes us from $(r_{\,1},r_{\,2})$ to $(r_+,r_-)$ (as in Fig.~\ref{fig:r1r2}):
\begin{eqnarray} \label{eq:rotation_r}
r_+ &\equiv& (r_{\,1} + r_{\,2})/\sqrt2 \nonumber \\
r_- &\equiv& (r_{\,2} - r_{\,1})/\sqrt2 \,.
\end{eqnarray}

In terms of the new variables, Eq.s~\ref{eq:param2} become
\begin{eqnarray}
\label{eq:E_+}
e_+ &=&  (y_{\,2} {+} y_{\,1})/\sqrt{2} - \theta\sqrt{2} - r_+ \\
\label{eq:E_-}
e_- &=& (y_{\,2} {-} y_{\,1})/\sqrt{2} - r_- \,.
\end{eqnarray}

We can form an argument similar to that in Sec.~\ref{sec:quantisation_error0}.
First we must establish the \pdf\ pertaining to $E_+$ after the measurement.
This \pdf\ will be conditioned not only by $y_{\,1}$ and $y_{\,2}$
but also by the unknown $r_+$;
that is, ${\Pr}_{E_+}(e_+|\{{\mathcal B}_{\,i}\},\{y_{\,i}\},r_+)$.
Each possible $r_+$
corresponds to a certain range of possible values for $r_-$,
say $[r_{-,a},r_{-,b})$ (as marked on Fig.~\ref{fig:r1r2});
from Eq.~\ref{eq:E_-} we obtain a corresponding range of values for $e_-$;
this range defines a strip in the $(e_+,e_-)$ plane;
the marginal \pdf\ of $e_+$ in this strip
(extracted from the joint \pdf\ of Eq.~\ref{eq:joint1})
is the \pdf\ of $E_+$ with reference to $y_{\,1}$, $y_{\,2}$, and $r_+$.

In defining this \pdf\ we have not required that $P_{\,i}$ be small
but, since this assumption will be needed soon,
we simplify the equations by introducing it from the beginning.
Therefore the strip in the plane $(e_+,e_-)$ will be reckoned
as approximately of infinitesimal width, with $e_-$ $\approx$ $0$.
In the following equation the limiting operation regards
an imagined sequence of pairs $(P_{\,1},P_{\,2})$
going to $(0,0)$ with a constant or bounded ratio.
The corresponding values $r_+$, each in
$\left[ - \left(P_{\,1}{+}P_{\,2}\right)/\left(2\sqrt{2}\right),
\left(P_{\,1}{+}P_{\,2}\right)/\left(2\sqrt{2}\right)\right)$,
make no difference in defining this limit:
\begin{equation} \label{eq:Pr_E+}
{\lim}~~{\Pr}_{E_+}(e_+|\{{\mathcal B}_{\,i}\},\{y_{\,i}\},r_+)
\; \propto \;
f_{\,1}(\frac{y_{\,2}{-}y_{\,1}}{2} {-} \frac{e_+}{\sqrt{2}}) \;
f_{\,2}(\frac{y_{\,2}{-}y_{\,1}}{2} {+} \frac{e_+}{\sqrt{2}}) \,.
\end{equation}
which serves in approximating the \pdf\ of $e_+$
up to a normalisation constant.

Now we can formulate predictive inference regarding $x$,
the datum of a future generic measurement,
which is modeled as in Eq.~\ref{eq:model}.
For any tetrad of tentative values $y_{\,1}$, $y_{\,2}$, $e_+$, and $r_+$,
using Eq.~\ref{eq:E_+} we find the corresponding value of the parameter,
\begin{equation}
\theta = (y_{\,2} {+} y_{\,1})/2 - e_+/\sqrt{2} - r_+/\sqrt{2} \,.
\end{equation}
In reference to the the tetrad $y_{\,1}$, $y_{\,2}$, $e_+$, and $r_+$
the future datum corresponds to a random variable:
\begin{equation}
{\Pr}_X(x|\{{\mathcal B}_{\,i}\},\{y_{\,i}\},e_+,r_+)
= p\left(x; \left(\left(y_{\,2} {+} y_{\,1}\right)/2-e_+/\sqrt{2}-r_+/\sqrt{2}
\right)\right) \,.
\end{equation}

$y_{\,1}$ and $y_{\,2}$ are known;
conditioned by $r_+$, a \pdf\ for $e_+$ is defined
and approximated in Eq.~\ref{eq:Pr_E+}, assuming that $P_{\,i}$ are small.
Therefore we define joint \pdf\ of $X$ and $E_+$
conditioned by $y_{\,1}$, $y_{\,2}$, and $r_+$,
which we express here in the limit of vanishing $P_{\,i}$,
\begin{eqnarray}
&{\lim}&
{\Pr}_{X,E_+}(x,e_+|\{{\mathcal B}_{\,i}\},{\mathcal M},\{y_{\,i}\},r_+)
\nonumber \\ &\propto&
f_{\,1}(\frac{y_{\,2}{-}y_{\,1}}{2} {-} \frac{e_+}{\sqrt{2}}) \;
f_{\,2}(\frac{y_{\,2}{-}y_{\,1}}{2} {+} \frac{e_+}{\sqrt{2}}) \;
p(x; \frac{y_{\,2}{+}y_{\,1}}{2}{-}\frac{e_+}{\sqrt{2}}) \,.
\end{eqnarray}
Predictive inference for $X$ amounts to the marginal \pdf\ of $X$;
in the limit of negligible $P_{\,i}$
\begin{eqnarray} \label{eq:pred_X-2}
&{\lim}&
{\Pr}_X(x|\{{\mathcal B}_{\,i}\},{\mathcal M},\{y_{\,i}\},r_+)
\nonumber \\ &\propto&
\int \, de_+ \; f_{\,1}(\frac{y_{\,2}{-}y_{\,1}}{2} {-}
\frac{e_+}{\sqrt{2}}) \;
f_{\,2}(\frac{y_{\,2}{-}y_{\,1}}{2} {+} \frac{e_+}{\sqrt{2}}) \;
p(x; \frac{y_{\,2}{+}y_{\,1}}{2}{-}\frac{e_+}{\sqrt{2}}) \,.
\end{eqnarray}
In this limit there is no dependence upon $r_+$.

As in the case of a single location measurement,
also from the combination of two location measurements
we have formulated predictive inference
conditioned on a function of the digitisation errors
with the understanding that, whatever that magnitude happens to be,
the corresponding true \pdf\ of $X$
would not deviate to a large extent from the density in Eq.~\ref{eq:pred_X-2}.

The same density function of $X$ can be obtained
if we pretend that
in the future measurement the parameter is a random variable
with \pdf\ $f_{\,1}(y_{\,1}{-}\theta)$ $f_{\,2}(y_{\,2}{-}\theta)$.
So this density function represents our prior-free assessment of $\theta$
based on the two location meassurements combined.
It is the posterior arising from a false prior uniform in $\theta$.

In terms of the discussion in Sec.~\ref{sec:naive}
(that is, disregarding the digitisation error)
we have defined probability density along line $l$,
defined by Eq.~\ref{eq:naive} on the $(E_{\,1},E_{\,2})$ plane,
simply as proportional to the pre-measurement joint \pdf\
of $E_{\,1}$ and $E_{\,2}$.

\section{Discussion}
Location measurements of the same parameter
are very easy to interpret intuitively,
so that the arguments in this work may be regarded as belabouring the obvious.
The reason for the systematic treatment is the need for preparing the concepts
pertaining to the study of generic measurements, even counting measurements,
perhaps also involving more than one parameters.
That development will be presented in another work.

We have examined in detail combinations
of only two location measurements of the same parameter.
Generalisation to any number $N$ of such measurements is straightforward.
After the measurements, the $N$-dimensional error will be constrained
close to a straight line by $N$$-$$1$ ancillary conditions
similar to Eq.~\ref{eq:E_-}.
In analogy to variables $E_-$ and $E_+$ of the two-measurement case,
$N$ new variables: $E_{-1}$, \dots $E_{-(N-1)}$, and $E_+$ are introduced.
In the same manner, $r_{-1}$, \dots $r_{-(N-1)}$, $r_+$ are defined.
The result is, in the approximation that assumes very small digitisation steps,
that predictive inference is approximately as if from a flat false prior.

Here is an idea for further research.
The simplicity of the result,
as expressed in the last paragraph of Sec.~\ref{sec:quantisation_error2},
perhaps indicates the existence of a proof more general
than the one presented in that section.
Such a proof would be advantageous if it could be generalised
to address the combination of location measurements
regarding not the same parameter but functions of the same parameter.
For example, combining a measurement of a side of a cube
with a measurement of the area of the cube and
with a measurement of the volume of the cube.
After the measurements,
the $N$-dimensional error is retricted on a curve
(if we disregard digitisation errors)
parametrized by $\theta$.
An obvious conjecture is that
the \pdf\ along this curve
will be proportional to the pre-measurement $N$-dimensional joint \pdf\ 
at the same point.
If we could prove it,
then also Jeffreys' prior would be shown to generate the same results.

\section{Conclusions}
Applying only the common postulates of probability,
without any specific assumptions about statistical inference,
from single or combined location measurements of the same parameter
we can obtain prior-free predictive inference.

The same density function would be obtained
if we started with a uniform prior for $\theta$.
Yet the logical standing of that Bayesian posterior
is distinct from prior-free assessment.
Moreover, we have not established 
either that in any generic situation predictive inference may be reduced
to a presumptive \pdf\ for the parameter,
or that such density function is the posterior from some false prior.

To be able to combine measurements, we must take into account
the inevitable digitisation errors in data recording, however small.
In the absence of prior probability for the parameter,
there are no exactly defined \pdfs\ for such errors.
Therefore we cannot just fold them into predictive inference;
that is, the true \pdf\ of predictive inference is unknown
but can be approximated by an available density function.

\section*{Acknowledgements}
I want to thank G\"unter Zech for some stimulating
exchanges that have inspired this work.
I also want to thank the Onassis Foundation
and the Ministry of Culture of Greece,
especially Mrs. Alcestis Soulogiannis.

\end{document}